\documentclass[a4paper,12pt]{revtex4}
\usepackage[latin1]{inputenc}
\usepackage{amsthm} 
\usepackage{amsmath}
\usepackage{graphicx} 
\usepackage{subfigure} 
\usepackage{amsfonts}
\usepackage{amssymb} 
\usepackage{hyperref} 
\usepackage{verbatim} 
\usepackage{pdflscape}
\usepackage{rotating}

\newtheorem{prop}{PROPOSITION}

\newtheorem{corol}{COROLLARY}

\begin{document}
\title{Equiangular tight frames and unistochastic matrices}

\author{Dardo Goyeneche}
\affiliation{Institute of Physics, Jagiellonian University, Krak\'ow, Poland}
\affiliation{Faculty of Applied Physics and Mathematics, Technical University of Gda\'{n}sk, 80-233 Gda\'{n}sk, Poland}
\affiliation{National Quantum Information Center in Gda\'{n}sk, 81-824 Sopot, Poland}

\author{Ond\v{r}ej Turek}

\affiliation{Nuclear Physics Institute, Academy of Sciences of the Czech Republic, 250 68 \v{R}e\v{z}, Czech Republic}
\affiliation{Bogolyubov Laboratory of Theoretical Physics, Joint Institute for Nuclear Research, 141980 Dubna, Russia}
\affiliation{Laboratory of Unified Quantum Devices, Kochi University of Technology, Kochi 782-8502, Japan}

\date{February 6, 2017}

\begin{abstract}
In this work, we show that a complex equiangular tight frame (ETF) composed by $N$ vectors in dimension $d$ exists if and only if a certain bistochastic matrix, univocally determined by $N$ and $d$, belongs to a special class of unistochastic matrices. This connection allows us to find new complex ETF in infinitely many dimensions and to derive a method to introduce non-trivial free parameters in ETF. We derive a 6-parametric family of complex ETF(6,16), which defines a family of symmetric POVM. Minimal and maximal possible average entanglement of the vectors within this qubit-qutrit family are presented. Furthermore, we propose an efficient numerical procedure to find the unitary matrix underlying a unistochastic matrix, which we apply to find all existing classes of complex ETF containing up to 19 vectors. 
\end{abstract}

\maketitle
 
\section{Introduction}\label{Introduction}
Equiangular tight frames (ETF) define regular geometrical structures in Hilbert spaces. An ETF composed of $N$ normalized vectors $\{\varphi_1,\dots,\varphi_N\}$ minimizes the highest possible coherence $\max_{j\neq k}|\langle\varphi_j|\varphi_k\rangle|$. These frames, which are also known as \emph{Grassmannian frames} \cite{SH03}, are thus typically related with optimal solutions of practical problems in communications, coding theory and sparse approximation \cite{JMF14,SH03,T04}. In particular, ETF provide an error-correcting code which is maximally robust against two erasures \cite{HP04}. ETF are also closely related to strongly regular graphs \cite{W09}, difference sets \cite{DF07,XZG74} and Steiner systems \cite{FMT12}. A special class of ETF consisting of $d^2$ vectors in $\mathbb{C}^d$ is known in quantum mechanics as \emph{Symmetric Informationally Complete Positive Operator Valued Measure} (SIC-POVM) \cite{RBSC04}. From considering rank-one SIC-POVM measurements it is possible to reconstruct any quantum state. Moreover, SIC-POVM are optimal measurements in the sense that the redundancy of information provided by any pair of outcomes of the POVM is minimized.

Along the years, many important results have been derived for real and complex ETF. A summary of the current state of the art can be found in a recent work of M. Fickus and D. Mixon \cite{FM15}.
However, the theory is still far from being complete. To illustrate this fact, let us mention that the full classification of complex ETF is only known in dimensions $d=2$ and $d=3$ \cite{S14}. The aim of the present work is to shed new light on ETF by establishing a connection with unistochastic matrices theory.
The paper is organised as follows: In Section \ref{S2} we review basic properties of equiangular tight frames. Section \ref{S3} is devoted to establishing the connection between ETF and unistochastic matrices. We study some fundamental properties and present new inequivalent classes of complex ETF in infinitely many dimensions. In Section \ref{S4} we present a method to introduce free parameters in a given real or complex ETF. In Section \ref{S5} we define an iterative procedure to find the unitary matrix underlying a unistochastic matrix. In Section \ref{S5} we summarize the results and pose some open questions. In Appendix \ref{Ap1} we explicitly derive a $6$-parametric family of complex ETF stemming from the real ETF($6,16$). Finally, Appendix \ref{Ap2} contains all the proofs of our propositions.

\section{Equiangular tight frames}\label{S2}
A complex ETF($d,N$) is a set of $N$ vectors $\{\varphi_k\}_{k=1,\dots,N}$ in $\mathbb{C}^d$ such that they are:
\begin{enumerate}
\item Normalized: $\|\varphi_j\|=1$ for every $j=1,\dots,N$.
\item Equiangular: $|\langle\varphi_j,\varphi_l\rangle|=\frac{1}{\alpha}$ for every $j\neq l=1,\dots,N$ and a fixed $\alpha>0$.
\item Tight frame: $\frac{d}{N}\sum_{j=1}^N\langle\varphi_j,\phi\rangle\varphi_j=\phi$ for every $\phi\in\mathbb{C}^d$.
\end{enumerate}
If the set of vectors forming the ETF belongs to $\mathbb{R}^d$, then we have a real ETF. It is easy to prove that the parameter $\alpha>0$ in property 2, which is called \emph{inverse coherence}, satisfies $\alpha=\sqrt{\frac{d(N-1)}{N-d}}$ (Welch bound). In quantum mechanics, the third property (tight frame) is equivalent to the fact that the rank-one projectors $\{\Pi_j=\langle\varphi_j,\cdot\rangle\,\varphi_j\}_{j=1,\dots,N}$ satisfy $\frac{d}{N}\sum_{j=1}^N\Pi_j=\mathbb{I}$, where $\mathbb{I}$ denotes the identity matrix, i.e., the projectors form a \emph{Positive Operator Valued Measure} (POVM) \cite{NC11}. 

The ETF problem can be equivalently posed in terms of Gram matrices $G_{jk}=\langle\varphi_j,\varphi_k\rangle$. A Gram matrix $G\in\mathbb{C}^{N\times N}$ is associated with an ETF($d,N$) if the following properties hold:
\begin{subequations}\label{Gram}
\begin{align}
G_{jj}&=1 \quad \text{for every $j=0,\dots N-1$}; \label{i} \\
|G_{jl}|&=\frac{1}{\alpha} \quad \text{for every $j\neq l=0,\dots N-1$ and a fixed $\alpha>0$}; \label{ii} \\
\sigma(G)&\in\{0,N/d\}, \label{iii}
\end{align}
\end{subequations}
where $\sigma(G)$ is the spectrum of $G$. It is easy to prove that the eigenvalues $N/d$ and $0$ of $G$ have multiplicities $d$ and $N-d$, respectively. When property \eqref{iii} is replaced by the weaker property $\mathrm{Rank}(G)=d$, the conditions \eqref{Gram} define a set of $N$ \emph{equiangular lines} in dimension $d$ \cite{LS73}. We recall that the vectors forming an ETF can be explicitly found from the Gram matrix by considering the \emph{Cholesky decomposition}, which can be efficiently implemented  (see \cite[page 52]{S13}).

\section{ETF and unistochastic matrices }\label{S3}
A bistochastic matrix $B$ is a square matrix of size $N$ having non-negative real entries such that 
\begin{equation}
\sum_{i=0}^{N-1}B_{ik}=1\hspace{0.3cm}\mbox{and}\hspace{0.3cm}\sum_{j=0}^{N-1}B_{kj}=1,
\end{equation}
for every $k=0,\dots N-1$. The matrix $B$ is called unistochastic if there exists a unitary matrix $U$ such that $B_{ij}=|U_{ij}|^2$, for every $i,j=0,\dots N-1$. If $U$ is a real orthogonal matrix then $B$ is called orthostochastic. Full set of unistochastic matrices is simple to derive for $N=2$ and it has been fully characterised also for $N=3$ \cite{BEKTZ}. The problem for $N=4$ is still open. A detailed explanation on unistochastic matrices and their applications can be found e.g. in Ref. \cite{S15}.

\begin{figure}[htbp]
\centering
\subfigure [Polar representation for ETF.]{\includegraphics[width=70mm]{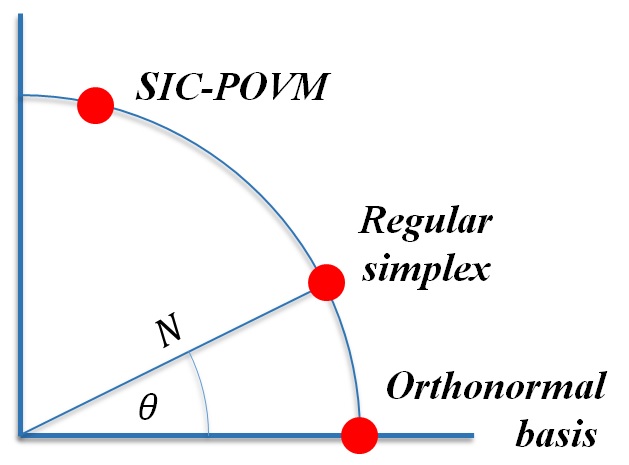}}\hspace{1cm}
\subfigure[ Existing ETF for $N=4,\dots,16$.]{\includegraphics[width=60mm]{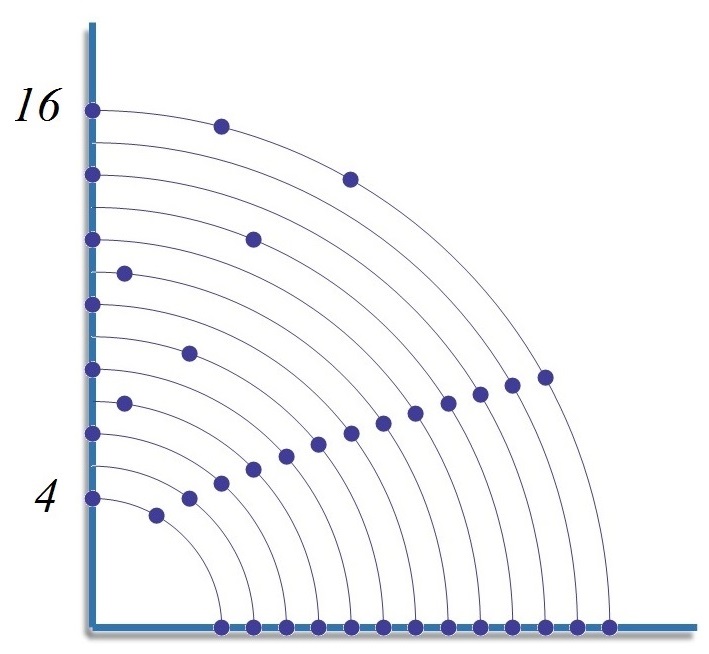}}
\caption{Polar representation for complex ETF induced by the existence of unistochastic matrices $\mathcal{B}_N(\theta)$. Proposition \ref{Prop1} allows us to identify coordinates $(\theta,N)$ unambiguously with ETF($d,N$), where $d=N\sin^2(\theta/2)$; E.g.,  orthonormal basis ($d=0,\,\theta=0$), regular simplices ($d=1,\,\sin(\theta/2)=1/\sqrt{N}$), and SIC-POVM ($d=\sqrt{N},\,\sin(\theta/2)=1/\sqrt{d}$). Note that $\theta\in[0,\pi/2]$ and $\theta\in[\pi/2,\pi]$ characterize ETF($d,N$) and ETF($N-d,N$) (Naimark complement), respectively. For example, $d=N$ and $d=N+1$ are the Naimark complements for orthonormal bases and regular simplices. Along this work we consider the convention $\theta\in[0,\pi/2]$, anti-clock wise.} 
\label{Fig1}
\end{figure}

From now on we will restrict our attention to a particular type of unistochastic matrices, denoted $\mathcal{B}_N(\theta)$. Namely, $\mathcal{B}_N(\theta)$ are unistochastic matrices such that there exists a unitary hermitian matrix $U_N(\theta)$ having a real and non-negative constant diagonal such that
\begin{equation}\label{bisto}
\mathcal{B}_N(\theta)_{ij}=|U_N(\theta)_{ij}|^2=\left\{\begin{array}{cl}
\cos^2(\theta) & \mbox{ if } i=j;\\
\frac{1}{N-1}\sin^2(\theta) & \mbox{ if } i\neq j,
\end{array}\right.
\end{equation}
where $\theta\in[0,\pi/2]$ for convenient reasons (see caption of Fig.\ref{Fig1}). We are now in position to establish the connection existing between ETF and unistochastic matrices $\mathcal{B}_N(\theta)$.
\begin{prop}\label{Prop1}
A complex ETF($d,N$) exists if and only if a unistochastic matrix $\mathcal{B}_N(\theta)$ exists for $d=N\sin^2(\theta/2)$, $\theta\in[0,\pi/2]$. In particular, the ETF($d,N$) is real if and only if $\mathcal{B}_N(\theta)$ is orthostochastic.
\end{prop}
Proofs of propositions can be found in Appendix \ref{Ap2}.
Unitary matrices having prescribed moduli of the entries have been previously studied \cite{AMM91, D94,ST08,TC15}. However, those approaches are only marginally related to our problem because hermiticity of the underlying unitary matrix and uniformity of the main diagonal are fundamental assumptions to connect ETF with unistochastic matrices.

Hermitian complex Hadamard matrices have a remarkable property, namely $U_{N_1}(\theta_{N_1})\otimes U_{N_2}(\theta_{N_2})=U_{N_1N_2}(\theta_{N_1N_2})$, where $\otimes$ denotes the Kronecker product. This property allows us to derive a wide range of inequivalent equiangular tight frames:
\begin{prop}\label{Prop2}
Let $\sqrt{N}=p_1^{r_1}\times\dots\times p_a^{r_a}$ be the prime power decomposition of $\sqrt{N}$, where $N$ is a square and $p_1,\dots,p_a$ are distinct prime numbers. Then there exist at least
\begin{equation}
\mathcal{N}=\mathcal{P}(r_1)\mathcal{P}(r_2)\dots \mathcal{P}(r_a),
\end{equation}
 inequivalent ETF($(N-\sqrt{N})/2,N$), where $\mathcal{P}(r)$ denotes the number of unrestricted partitions of the integer number $r$.
\end{prop}
Our approach based on unistochastic matrices enables us to address another interesting problem in the ETF theory, namely, the possibility to have the off-diagonal entries in the Gram matrix proportional to roots of unity. In Proposition \ref{Prop3} below, we show that for a given value of $N$ only certain $m$-th roots of unity are allowed. 
\begin{prop}\label{Prop3}
Consider an ETF with a Gram matrix such that its off-diagonal entries, when normalized to have absolute value $1$, are $m$-th roots of unity. Let $m=p_1^{a_1}\dots p_r^{a_r}$ be the decomposition of $m$ in prime power factors. Then there exist numbers $x_j\in\mathbb{N}\cup\{0\}$ such that $2k+N-2=x_1p_1+\cdots+x_rp_r$, where $k=\cot(\theta)\sqrt{N-1}$. 
\end{prop}
Using relation $d=N\sin^2(\theta/2)$, derived in Proposition \ref{Prop1}, one can express $\cot(\theta)$ in terms of $N$ and $d$. Hence we obtain
\begin{equation}\label{k}
k=\frac{N-2d}{2}\sqrt{\frac{N-1}{d(N-d)}}.
\end{equation}
Proposition \ref{Prop3} implies in particular that if $2k$ is not an integer number, then the normalized off-diagonal entries of the Gram matrix cannot be roots of unity. Note that for a real ETF the number $k$ is always integer \cite{STDH07}.

In order to illustrate the result, let us consider $N=64$ and fourth roots of unity. According to Proposition \ref{Prop3}, for $m=2^2$
we have $2k+64-2=2x_1$. The only ETF compatible with this equation and equation \eqref{k} are ETF(32,64), ETF(28,64) and ETF(8,64). Let us note that ETF(8,64) is a special kind of SIC-POVM known as \emph{Hoggar lines} \cite{H98}. In general, the following restriction holds for SIC-POVM.
\begin{corol}\label{Corol1}
A SIC-POVM in dimension $d$ admits a Gram matrix with off-diagonal entries proportional to roots of unity if $d=2$ or $d+1$ is a square.
\end{corol} 
The proof is staightforward from the fact that $2k$ has to be integer (see Prop. \ref{Prop3}) combined with equation \eqref{k}. Concerning the existence of SIC-POVM having Gram matrix composed by roots of unity up to dimension $d=100$, we have:
\begin{itemize}
\item $d=2,\hspace{0.1cm}3,\hspace{0.1cm} 8$: the existence of SIC-POVM confirmed with 4th, 6th and 4th roots of unity in the off-diagonal entries of the Gram matrix, respectively;
\item $d=15,\hspace{0.1cm} 24,\hspace{0.1cm} 35,\hspace{0.1cm} 48$: SIC-POVM exist, but the possibility of having roots of unity in the off-diagonal entries is open;
\item $d=63,\hspace{0.1cm} 80,\hspace{0.1cm} 99$: the existence of SIC-POVM is open.
\end{itemize}

\section{Free parameters for ETF}\label{S4}
In this section, we present a simple method to construct a parametric family of complex ETF stemming from a given ETF in dimension $d=(N-\sqrt{N})/2$. The construction is based on the following idea, introduced in Ref. \cite{G13}: Two vectors $v,w\in\mathbb{C}^N$ are called \emph{equivalent to real (ER) pair} if $v\circ w^*\in\mathbb{R}^N$, where the circle and asterisk denote the Hadamard (entrywise) product and complex conjugation, respectively. Let $C_A$ and $C_B$ be two columns of a complex Hadamard matrix that form an ER pair. Note that the entries of the vector $C_A\circ C_B^*$ are $\pm1$. We introduce a parameter $\alpha\in[0,2\pi)$ in vectors $C_A(\alpha)$ and $C_B(\alpha)$ as follows:
\begin{itemize}
\item If $(C_A\circ C_B^*)_j=-1$, we set $(C_A(\alpha))_j=e^{i\alpha}(C_A)_j$ and $(C_B(\alpha))_j=e^{i\alpha}(C_B)_j$;
\item if $(C_A\circ C_B^*)_j=1$, we set $(C_A(\alpha))_j=(C_A)_j$ and $(C_B(\alpha))_j=(C_B)_j$.
\end{itemize}
Columns $C_A(\alpha)$ and $C_B(\alpha)$ are orthogonal for any $\alpha\in[0,2\pi)$ and they belong to the plane defined by $C_A(0)$ and $C_B(0)$, i.e., they are orthogonal to the rest of the columns of the Hadamard matrix. Therefore, if we replace columns $C_A$ and $C_B$ in the original Hadamard matrix with $C_A(\alpha)$ and $C_B(\alpha)$, respectively, we obtain a family of complex Hadamard matrices parametrized by $\alpha$.
Note that real pairs of columns form trivially ER pairs, thus they always allow to introduce a free parameter.

In order to illustrate the procedure, let us consider the Hadamard matrix
$$
H=\left(\begin{array}{rrrr}
-1&1&1&1\\
1&-1&1&1\\
1&1&-1&1\\
1&1&1&-1
\end{array}\right).
$$
Let $C_A=(1,1,-1,1)^T$ and $C_B=(1,1,1,-1)^T$ be the third and the fourth column of $H$. Since $C_A\circ C_B^*=(1,1,-1,-1)^T$, we set $C_A(\alpha)=(1,1,-e^{i\alpha},e^{i\alpha})^T$ and $C_B(\alpha)=(1,1,e^{i\alpha},-e^{i\alpha})^T$. Then
\begin{equation}\label{Ht}
H(\alpha)=\left(\begin{array}{rrrr}
-1&1&1&1\\
1&-1&1&1\\
1&1&-e^{i\alpha}&e^{i\alpha}\\
1&1&e^{i\alpha}&-e^{i\alpha}
\end{array}\right)
\end{equation}
is a 1-parametric family of complex Hadamard matrices.

\begin{figure}[!h]
\centering 
{\includegraphics[width=8cm]{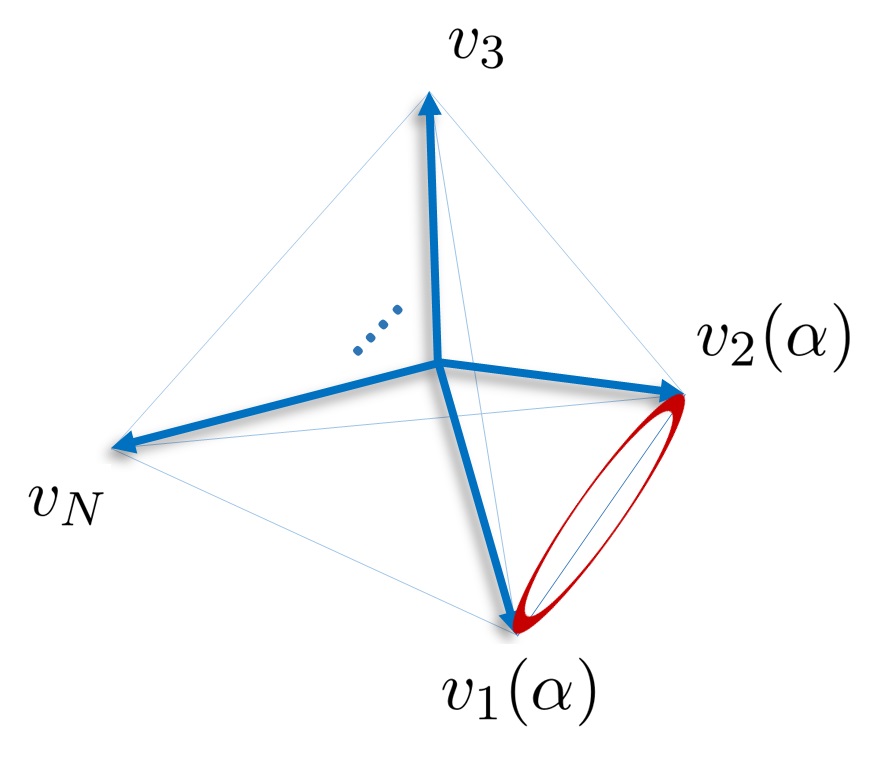}} 
\caption{Introduction of a free parameter $\alpha$ in a pair of vectors of an ETF. A complex cone is generated by vectors $v_1(\alpha)$ and $v_2(\alpha)$ in such a way that the $N$ vectors $\{v_1(\alpha),v_2(\alpha),v_3,\dots,v_N\}$ define a complex ETF for every $\alpha\in[0,\pi)$. Note that $v_2(\pi)=v_1(0)$ and $v_1(\pi)=v_2(0)$.}
\label{Fig2}
\end{figure}

As we have mentioned in Section~\ref{S3}, real and complex Hadamard matrices of size $N$ are associated with real and complex ETF$((N-\sqrt{N})/2,N)$, respectively. The technique of introducing free parameters in a Hadamard matrix can be thus used for constructing parametric families of ETF$((N-\sqrt{N})/2,N)$, cf. Fig. \ref{Fig2}. It is important to remark that our method has to be applied to both columns and rows in order to obtain a Hermitian matrix. That is, the ER pair of columns $\{C_A(\alpha),C_B(\alpha)\}$ has to be complemented with the associated ER pair of rows $\{R_A(-\alpha),R_B(-\alpha)\}$, where the parameter $\alpha$ is reflected due to the requirement of hermiticity. Let us add some clarifying remarks:
\begin{itemize}
\item In the special case $N=4$, a hermitian family of complex Hadamard matrices cannot be constructed from a real symmetric Hadamard matrix with constant diagonal, because introducing rows $\{R_A(-\alpha),R_B(-\alpha)\}$ in addition to columns $\{C_A(\alpha),C_B(\alpha)\}$ in Eq.(\ref{Ht}) leads to the matrix $H(0)$. Equivalently, a family of ETF(3,4) does not exist.
\item Hermitian families exist for every real symmetric Hadamard matrix having constant diagonal 1 of even square size $N>4$, as ER pairs always exist for such cases. The first hermitian family stemming from a real Hadamard occurs for $N=16$. The 6-parametric family is presented in Appendix \ref{Ap1}. It is worth noting that this method allows us to introduce up to $N/2-1$ free parameters.
\item We can also introduce $N/2-1$ free parameters for hermitian Fourier matrices when $N$ is even \cite{G13}.  For other hermitian complex Hadamard matrices of even square size $N$ the method cannot be applied for the reason that ER pairs do not exist.
\end{itemize}
Now we are ready to formulate the main result of this section, which is based on the considerations above.
\begin{prop}\label{Prop5}
There exists complex ETF$((N-\sqrt{N})/2,N)$ admitting the introduction of $N/2-1$ linearly independent free parameters for every even square value of $N>4$. For real ETF$((N-\sqrt{N})/2,N)$ the result holds if and only if a real symmetric Hadamard matrix exists.
\end{prop}
Proof follows from Theorem 3.1 in Ref.\cite{G13}. For odd values of $N$, free parameters cannot be introduced by using our method. This is so because ER pairs do not exist if $N$ is odd \cite{G13}. Given that ER pairs is the most general method to introduce free parameters in pair of columns (see \cite{G13}) we conclude that the rotation of two vectors of any ETF$((N-\sqrt{N})/2,N)$ cannot generate a family of ETF for any odd $N$.

\section{Entanglement in equiangular tight frames}\label{S5}
Equiangular tight frames have important applications in quantum mechanics, as they represent symmetric POVM quantum measurements. It is therefore interesting to study entanglement of the symmetric vectors forming a POVM, which could be important for some theoretical and experimental implementation purposes. In this section we discuss average entanglement properties of both SIC-POVM and the 6-parametric family of ETF(6,16), defined in Appendix \ref{Ap1}. 

Before proceeding forward let us recall that purity of a quantum state $\rho$ is given by $P=\mathrm{Tr}(\rho^2)$. It quantifies how close is a quantum state to the surface of pure states, i.e, the surface of rank one-projectors: if $\rho$ is defined in dimension $d$, then $1/d\leq P\leq 1$, where $P=1/d$ holds for the maximally mixed state $\rho=\mathbb{I}/d$ and $P=1$ for any pure state $\rho=|\phi\rangle\langle\phi|$. For example, the entire set of quantum states for qubit systems satisfying $P=1$ determine the surface of the Bloch sphere \cite{NC11} (Poincar\'{e} sphere for mathematicians), whereas the center of the sphere represents the maximally mixed state.

An ETF($d,d^2$), i.e. SIC-POVM in dimension $d$, defined for a bipartite system $d=d_A\times d_B$ has a fixed average purity of reductions \cite{ZTE10}:
\begin{equation}\label{avpuSIC}
\frac{1}{d_Ad_B}\sum_{j=1}^{d_Ad_B}\mathrm{Tr}(\rho^2_{A_j})=(d_A + d_B)/(d_Ad_B + 1).
\end{equation}
Here, $\rho_{A_j}=\mathrm{Tr}_B(|\psi_j\rangle\langle\psi_j|)$ is the reduction to the first subsystem $A$ and $|\psi_j\rangle$ is the $j$th element of the SIC-POVM. The symbol $\mathrm{Tr}_B$ means partial trace over the second subsystem $B$. Eq.(\ref{avpuSIC}) also holds for the average purity of reductions to subsystem $B$, as the bipartite states forming the POVM are pure. The key property required to derive Eq.(\ref{avpuSIC}) is the fact that SIC-POVM are 2-designs. Other classes of ETF are only 1-designs and, therefore, it is expected that similar results do not hold. 

As a novel contribution, here we show that the average purity of reductions for the 6-parametric family of ETF(6,16), presented in Appendix \ref{Ap1} depends on the parameters, and we present approximate lower and upper bounds. This family of symmetric POVM is defined for a qubit-qutrit system ($d=6=d_A\times d_B=2\times 3$). The average purity for the qubit subsystem (A) seems to lie in the range
\begin{equation}\label{bounds}
0.576737\lessapprox\frac{1}{16}\sum_{j=1}^{16}\mathrm{Tr}(\rho^2_{A_j})\lessapprox0.804885,
\end{equation}
where the approximate lower (LB) and upper (UB) bounds are attained for parameters
\begin{equation}
\vec{\alpha}_{LB}=\{0.0970, 0.0957, 0.4536, 0.7275, 0.7287, 0.2258\},
\end{equation}
and
\begin{equation}
\vec{\alpha}_{UB}=\{2.2222,2.2233,3.1401,0.4173,2.9043,2.6317\},
\end{equation}
respectively. The approximate bounds given in Eq.(\ref{bounds}) were obtained from considering numerical optimizations in Mathematica. An analytic derivation of the bounds seems to be a hard problem. As an interesting consequence of Eq.(\ref{bounds}), we noted that the 16 vectors forming the ETF(6,16) can be neither fully separable nor maximally entangled, because the average purity of reductions can attain neither the value $UB=1$ nor the value $LB=1/2$. 

In general, for $N$ vectors defining an ETF in dimension $d\gg \sqrt{N}$ it is reasonable to expect weak restrictions on the average purity of reductions. Indeed, vectors forming an ETF($d,d$), i.e. an orthonormal basis in dimension $d$, can be fully separable (e.g., tensor product basis in dimension $d=2^N$ for an $N$ qubit system) or maximally entangled (e.g. generalized Bell basis in dimension $d=d'\times d'$ for a two-qudit system having $d'$ internal levels each).

\section{Algorithm to find the underlying unitary matrix}\label{S6}

In this section we present an iterative algorithm that efficiently finds the underlying unitary matrix $U$ existing behind a unistochastic matrix $B$. The algorithm works as follows:

1. [Seed]: Start from a random matrix $A^{(0)}$ of size $N$.

2. [Bistochasticity]: $A^{(0)}_{i,j}\,\rightarrow\,A^{(1)}_{i,j}=\frac{A_{i,j}}{|A_{i,j}|}\sqrt{B_{i,j}}$.

3. [Unistochasticity]: Apply the Schmidt orthogonalization to columns of $A^{(1)}$.

The iteration of steps 2 and 3 generate a sequence of matrices that converges to a unitary matrix $U$ such that $B_{ij}=|U_{ij}|^2$, for a suitable choice of the seed in step 1. Indeed, unistochastic matrices are attractive fixed points of the composed non-linear operators $T=T_3T_2$, where $T_2$ and $T_3$ are the non-linear operators associated to the mappings defined in steps 2 and 3, respectively. A solution is found when the seed belongs to the basin of attraction of a unistochastic matrix $B$. The additional steps: 

$2^{\prime}$. \, [Hermiticity]: $A\,\rightarrow\,(A+A^{\dag})/2$,
 
$2^{\prime\prime}$. \, [Constant diagonal]: $A_{i,i}\,\rightarrow \cos(\theta)$, where $d=N\sin^2(\theta/2)$, \vspace{0.2cm}

\noindent generate the unitary hermitian matrix $U_N(\theta)$ having constant diagonal and, consequently, the unistochastic matrix $\mathcal{B}_N(\theta)$, where $d=N\sin^2(\theta/2)$. In this way, we determine a complex ETF($d,N$). A similar procedure can be applied to generate a real ETF. In the case that the imposed matrix $B$ is bistochastic but not unistochastic, the algorithm exhibit oscillations without converging. 

We have implemented the above described procedure for matrices of size $N=4,\dots,22$. Up to dimension $d=19$ we were able to reproduce all known classes of complex ETF \cite{FM15} by considering $10^3$ seeds. This number of seeds was no more sufficient in dimension $d=20$, where  $10^4$ seeds were required to find ETF(10,20).

In dimension $d=22$ we focused our attention on the existence of the complex ETF(11,22), which is still open. Such case would have associated an hermitian complex conference matrix $U_{22}(\pi/2)$. Our simulations considered $10^6$ random seeds and we could not find a single successful convergence of our iterative procedure, which suggests that  an hermitian complex conference matrix of size 22 does not exist. On the other hand, note that a real symmetric conference matrix of size 22 does not exist, which excludes the existence of a real ETF(11,22). This is so because the number 22 is not the sum of two integer squares \cite{B50}.

\section{Conclusions}\label{S7}

We have introduced a one-to-one connection between complex equiangular tight frames (ETF) and a special kind of unistochastic matrices $\mathcal{B}_N(\theta)$, defined in Eq.(\ref{bisto}). The connection has been established in Prop. \ref{Prop1}. As a direct consequence, we have found new classes of complex ETF (see Prop. \ref{Prop2}). Furthermore, we presented new integrality restrictions for some classes of  equiangular tight frames, i.e., for those having Gram matrix whose off-diagonal entries are proportional to roots of unity (see Prop. \ref{Prop3}). In particular, the only possible SIC-POVM of such kind may only exist in dimension $d=2$ or when $d+1$ is a square number (see Corol. \ref{Corol1}). We also proposed a method to introduce non-trivial free parameters in real and complex ETF (see Prop. \ref{Prop5}). To illustrate our method, we explicitly derived a 6-parametric family of complex ETF stemming from the real ETF(6,16) (see Appendix \ref{Ap1}). Moreover, we studied the average purity of reductions ($P$) for this  family, which defines a symmetric POVM for a qubit-qutrit quantum system. As consequence, we have found that $P$ can be neither maximal ($P=1$ for separable states) nor minimal ($P=1/2$ for maximally entangled states). Lower and upper bounds for $P$ were derived (see Section \ref{S6}).

Furthermore, we presented an efficient algorithm to find unistochastic matrices, which is simple to implement in a computer language. By introducing a refinement of this procedure, we can also find the unitary hermitian matrices relying behind unistochastic matrices $\mathcal{B}_N(\theta)$, which allows to find an ETF($d,N$), where $d=N\sin^2(\theta/2)$. By using this procedure, we calculated all the parameters ($d,N$) for which a complex ETF($d,N$) exists, up to $N=19$. The results are consistent with known classes described in the literature \cite{FM15}. Additionally, we exhaustively studied the existence of the complex ETF(11,22), which seems not to exist. This result would imply that an hermitian complex conference matrix of size 22 does not exist.

We conclude the paper with two important open questions: (\emph{i}) Find integrality restrictions for a general complex ETF, and (\emph{ii}) Solve the Fickus conjecture \footnote{See the blog \emph{Short, fat matrices} by D. Mixon,  https://dustingmixon.wordpress.com/2015/07/08/conjectures-from-sampta/}: Consider the three numbers $d$, $N-1$ and $N-d$. If a complex ETF($d,N$) exists then one of these three numbers divides the product of the other two. Our numerical simulations up to $N=19$, as well as all solutions presented in the most complete catalog of ETF, Ref. \cite{FM15}, are consistent with the conjecture.

\section*{Acknowledgements}
DG thanks to S. Friedline, M. Fickus and K. \.{Z}yczkowski for fruitful discussions on equiangular tight frames and unistochastic matrices. This work has been supported by the Polish National Science Center under the project number DEC-2011/02/A/ST1/00119 and by the Czech Science Foundation (GA\v{C}R) within the project 17-01706S.

\appendix

\section{ETF(6,16) having 6 free parameters}\label{Ap1}
In this section we explicitly present a 6-parametric family of hermitian unitary matrices $U(\vec{\alpha})$ that determines the existence of a family of complex ETF(6,16). This family stems from the real symmetric Hadamard matrix $U(\vec{0})=H_2^{\otimes 4}$, where $\otimes$ denotes the Kronecker product. The existing ER pairs of columns and rows are given by: $\{2,3\};\{4,13\};\{5,8\};\{6,7\};\{9,12\};\{10,11\};\{14,15\}$, which are associated to the free parameters $\alpha_1$ to $\alpha_7$, respectively. Therefore, the hermitian unitary matrix containing 6 free parameters is given by
\newpage
\scalebox{0.85}{
\begin{sideways}
\begin{minipage}{\textheight}
\vspace{2cm}
\begin{equation}
\hspace{-7cm}U_{16}(\vec{\alpha})=\frac{1}{4}\left(\begin{array}{cccccccccccccccc}
1&1&1&1&1&1&1&1&1&1&1&1&1&1&1&1\\
1&1&-1&-1&e^{i(\alpha_2-\alpha_1)}&e^{i(\alpha_2-\alpha_1)}&-e^{i(\alpha_2-\alpha_1)}&-e^{i(\alpha_2-\alpha_1)}&1&1&-1&-1&e^{i(\alpha_2-\alpha_1)}&e^{i(\alpha_2-\alpha_1)}&-e^{i(\alpha_2-\alpha_1)}&-e^{i(\alpha_2-\alpha_1)}\\
1&-1&1&-1&e^{i(\alpha_3-\alpha_1)}&-e^{i(\alpha_3-\alpha_1)}&e^{i(\alpha_3-\alpha_1)}&-e^{i(\alpha_3-\alpha_1)}&1&-1&1&-1&e^{i(\alpha_3-\alpha_1)}&-e^{i(\alpha_3-\alpha_1)}&e^{i(\alpha_3-\alpha_1)}&-e^{i(\alpha_3-\alpha_1)}\\
1&-1&-1&1&e^{-i\alpha_1}&-e^{-i\alpha_1}&-e^{-i\alpha_1}&e^{-i\alpha_1}&1&-1&-1&1&e^{-i\alpha_1}&-e^{-i\alpha_1}&-e^{-i\alpha_1}&e^{-i\alpha_1}\\
1&e^{i(\alpha_1-\alpha_2)}&e^{i(\alpha_1-\alpha_3)}&e^{i\alpha_1}&1&e^{-i(\alpha_4-\alpha_5)}&e^{-i(\alpha_4-\alpha_6)}&e^{-i\alpha_4}&-1&-e^{i(\alpha_1-\alpha_2)}&-e^{i(\alpha_1-\alpha_3)}&-e^{i\alpha_1}&-1&-e^{-i(\alpha_4-\alpha_5)}&-e^{-i(\alpha_4-\alpha_6)}&-e^{-i\alpha_4}\\
1&e^{i(\alpha_1-\alpha_2)}&-e^{i(\alpha_1-\alpha_3)}&-e^{i\alpha_1}&e^{i(\alpha_4-\alpha_5)}&1&-e^{-i(\alpha_5-\alpha_6)}&-e^{-i\alpha_5}&-1&-e^{i(\alpha_1-\alpha_2)}&e^{i(\alpha_1-\alpha_3)}&e^{i\alpha_1}&-e^{i(\alpha_4-\alpha_5)}&-1&e^{-i(\alpha_5-\alpha_6)}&e^{-i\alpha_5}\\
1&-e^{i(\alpha_1-\alpha_2)}&e^{i(\alpha_1-\alpha_3)}&-e^{i\alpha_1}&e^{i(\alpha_4-\alpha_6)}&-e^{i(\alpha_5-\alpha_6)}&1&-e^{-i\alpha_6}&-1&e^{i(\alpha_1-\alpha_2)}&-e^{i(\alpha_1-\alpha_3)}&e^{i\alpha_1}&-e^{i(\alpha_4-\alpha_6)}&e^{i(\alpha_5-\alpha_6)}&-1&e^{-i\alpha_6}\\
1&-e^{i(\alpha_1-\alpha_2)}&-e^{i(\alpha_1-\alpha_3)}&e^{i\alpha_1}&e^{i\alpha_4}&-e^{i\alpha_5}&-e^{i\alpha_6}&1&-1&e^{i(\alpha_1-\alpha_2)}&e^{i(\alpha_1-\alpha_3)}&-e^{i\alpha_1}&-e^{i\alpha_4}&e^{i\alpha_5}&e^{i\alpha_6}&-1\\
1&1&1&1&-1&-1&-1&-1&1&1&1&1&-1&-1&-1&-1\\
1&1&-1&-1&-e^{i(\alpha_2-\alpha_1)}&-e^{i(\alpha_2-\alpha_1)}&e^{i(\alpha_2-\alpha_1)}&e^{i(\alpha_2-\alpha_1)}&1&1&-1&-1&-e^{i(\alpha_2-\alpha_1)}&-e^{i(\alpha_2-\alpha_1)}&e^{i(\alpha_2-\alpha_1)}&e^{i(\alpha_2-\alpha_1)}\\
1&-1&1&-1&-e^{i(\alpha_3-\alpha_1)}&e^{i(\alpha_3-\alpha_1)}&-e^{i(\alpha_3-\alpha_1)}&e^{i(\alpha_3-\alpha_1)}&1&-1&1&-1&-e^{i(\alpha_3-\alpha_1)}&e^{i(\alpha_3-\alpha_1)}&-e^{i(\alpha_3-\alpha_1)}&e^{i(\alpha_3-\alpha_1)}\\
1&-1&-1&1&-e^{-i\alpha_1}&e^{-i\alpha_1}&e^{-i\alpha_1}&-e^{-i\alpha_1}&1&-1&-1&1&-e^{-i\alpha_1}&e^{-i\alpha_1}&e^{-i\alpha_1}&-e^{-i\alpha_1}\\
1&e^{i(\alpha_1-\alpha_2)}&e^{i(\alpha_1-\alpha_3)}&e^{i\alpha_1}&-1&-e^{-i(\alpha_4-\alpha_5)}&-e^{-i(\alpha_4-\alpha_6)}&-e^{-i\alpha_4}&-1&-e^{i(\alpha_1-\alpha_2)}&-e^{i(\alpha_1-\alpha_3)}&-e^{i\alpha_1}&1&e^{-i(\alpha_4-\alpha_5)}&e^{-i(\alpha_4-\alpha_6)}&e^{-i\alpha_4}\\
1&e^{i(\alpha_1-\alpha_2)}&-e^{i(\alpha_1-\alpha_3)}&-e^{i\alpha_1}&-e^{i(\alpha_4-\alpha_5)}&-1&e^{-i(\alpha_5-\alpha_6)}&e^{-i\alpha_5}&-1&-e^{i(\alpha_1-\alpha_2)}&e^{i(\alpha_1-\alpha_3)}&e^{i\alpha_1}&e^{i(\alpha_4-\alpha_5)}&1&-e^{-i(\alpha_5-\alpha_6)}&-e^{-i\alpha_5}\\
1&-e^{i(\alpha_1-\alpha_2)}&e^{i(\alpha_1-\alpha_3)}&-e^{i\alpha_1}&-e^{i(\alpha_4-\alpha_6)}&e^{i(\alpha_5-\alpha_6)}&-1&e^{-i\alpha_6}&-1&e^{i(\alpha_1-\alpha_2)}&-e^{i(\alpha_1-\alpha_3)}&e^{i\alpha_1}&e^{i(\alpha_4-\alpha_6)}&-e^{i(\alpha_5-\alpha_6)}&1&-e^{-i\alpha_6}\\
1&-e^{i(\alpha_1-\alpha_2)}&-e^{i(\alpha_1-\alpha_3)}&e^{i\alpha_1}&-e^{i\alpha_4}&e^{i\alpha_5}&e^{i\alpha_6}&-1&-1&e^{i(\alpha_1-\alpha_2)}&e^{i(\alpha_1-\alpha_3)}&-e^{i\alpha_1}&e^{i\alpha_4}&-e^{i\alpha_5}&-e^{i\alpha_6}&1
\end{array}\right).
\end{equation}
\end{minipage}
\end{sideways}}
\newpage

The explicit form of the 16 vectors can be found by considering the Cholesky or singular value decomposition of the Gram matrix $G=4/3(\mathbb{I}-U_{16}(\vec{\alpha}))$.

\section{Proof of propositions}\label{Ap2}

\noindent\textbf{Proposition \ref{Prop1}:}
A complex ETF($d,N$) exists if and only if a unistochastic matrix $\mathcal{B}(\theta)$ exists for $d=N\sin^2(\theta/2)$. In particular, the ETF($d,N$) is real if and only if $\mathcal{B}(\theta)$ is orthostochastic.
\begin{proof}
Suppose a complex ETF($d,N$) exists, having associated a Gram matrix $G$. Therefore, $U=\mathbb{I}-\frac{2d}{N}G$ is a unitary hermitian matrix that implies the existence of the unistochastic matrix $\mathcal{B}(\theta)$ defined in Eq.(\ref{bisto}). Furthermore, by using the fact that $U_{ii}=1-2d/N$ and Eq.(\ref{bisto}) we have $N\cos(\theta)=N-2d$, which is equivalent to $d=N\sin^2(\theta/2)$. Reciprocally, suppose that there exists a unistochastic matrix $\mathcal{B}(\theta)$ with an underlying hermitian unitary matrix $U_N(\theta)$. By checking properties \eqref{Gram}, let us prove that $G=\frac{N}{2d}(\mathbb{I}-U_N(\theta))$ for $d=N\sin^2(\theta/2)$ is a Gram matrix of an ETF($d,N$). As for property \eqref{i}, we have $G_{ii}=\frac{N}{2d}(1-\cos(\theta))=\frac{N}{d}\sin^2(\theta/2)=1$, where we used the fact that $U_{ii}=\cos(\theta)$. Similarly, we prove \eqref{ii}, i.e., $|G_{ij}|^2=\frac{N-d}{d(N-1)}$ for $i\neq j$ (Welch bound). As for \eqref{iii}, since $U_N(\theta)$ is hermitian unitary with constant diagonal, $U_N(\theta)$ has eigenvalues $1$ and $-1$ and $\mathrm{Tr}(G)=N$; hence $G$ has eigenvalues $\lambda_0=0$ and eigenvalues $\lambda_1=\frac{N}{d}$ with degeneracies $N-d$ and $d$, respectively. To sum up, $G$ is a Gram matrix of a complex ETF($d,N$). In particular, real matrices $G$ associated to real ETF($d,N$) are one-to-one related to orthostochastic matrices $\mathcal{B}_N(\theta)$, where $U_N(\theta)$ is an orthogonal symmetric matrix.
\end{proof}

\noindent\textbf{Proposition \ref{Prop2}:}
Let $\sqrt{N}=p_1^{r_1}\times\dots\times p_a^{r_a}$ be the prime power decomposition of $\sqrt{N}$, where $N$ is a square and $p_1,\dots,p_a$ are distinct prime numbers. Then there exist at least
\begin{equation}
\mathcal{N}=\mathcal{P}(r_1)\mathcal{P}(r_2)\dots \mathcal{P}(r_a),
\end{equation}
 inequivalent ETF($(N-\sqrt{N})/2,N$), where $\mathcal{P}(r)$ denotes the number of unrestricted partitions of the integer number $r$.
\begin{proof}
Let $\sqrt{N}=p_1^{r_1}\times\dots\times p_a^{r_a}$ be the prime power decomposition of $\sqrt{N}$. Given that hermitian Fourier matrices exist in every square dimension (see Ref. \cite{BE10,Sol13}) we can construct matrices $U(\theta)$ with $\cos(\theta)=1/\sqrt{N}$ by considering tensor product of hermitian Fourier matrices of sizes $p_1^{2s_1},p_2^{2s_2},\dots,p_a^{2s_a}$ for every $s_j=1,\dots,r_j$. In particular, all possible products of matrices of size $p_1^{2s_1}$ generate $\mathcal{P}(r_1)$ different matrices of size $p_1^{2r_1}$. Therefore, there are $\mathcal{N}=\mathcal{P}(r_1)\mathcal{P}(r_2)\dots \mathcal{P}(r_a)$ different ways to construct a matrix $U(\theta)$ of size $N=p_1^{2r_1}\times\dots\times p_a^{2r_a}$. Furthermore, all of these matrices are inequivalent in the sense that we cannot transform one into the other by permuting rows/columns and multiplying rows/columns by unimodular complex numbers. This is so because the set of invariants under equivalence known as \emph{Haagerup set} \cite{H96}:
\begin{equation}
\Lambda(U)=\{U_{st}U_{uv}U^*_{sv}U^*_{ut}\,:\,s,t,u,v=0,\dots,N-1\},
\end{equation}
is different for all the constructed matrices $U(\theta)$. Therefore, the ETF obtained from them are inequivalent.
\end{proof}
To exemplify unrestricted partitions, note that $4=4+0=3+1=2+2=2+1+1=1+1+1+1$, so $\mathcal{P}(4)=5$. Proposition \ref{Prop2} can be extended to a construction of other ETF($(N-\sqrt{N})/2,N$) by considering some existing real 1-weighing (Hadamard) matrices. For example, for $N=144$ the real case $4\times36$ is not considered in Proposition(\ref{Prop2}). Also, there are further complex cases such as $4\times 36$, generated by considering the tensor product of the symmetric Hadamard matrix of size 4 and the hermitian Fourier matrix of size 36; and so on for higher dimensions.\vspace{0.3cm}

\noindent\textbf{Proposition \ref{Prop3}:}
Consider an ETF with a Gram matrix such that its off-diagonal entries, when normalized to have absolute value $1$, are $m$-th roots of unity. Let $m=p_1^{a_1}\dots p_r^{a_r}$ be the decomposition of $m$ in prime power factors. Then there exist numbers $x_j\in\mathbb{N}\cup\{0\}$ such that $2k+N-2=x_1p_1+\cdots+x_rp_r$, where $k=\cot(\theta)\sqrt{N-1}$. 
\begin{proof}
As we have seen in the proof of Proposition \ref{Prop1}, the off-diagonal entries of the Gram matrix coincide with the off-diagonal entries of the associated matrix $U(\theta)$. We multiply the columns of $U(\theta)$ so that the first two rows have the form
\begin{eqnarray}
\{\cos(\theta),1,\dots,1\}\hspace{0.5cm}\mbox{and}\hspace{0.5cm}\{1,\cos(\theta),e^{i\alpha_1}\sin(\theta)/\sqrt{N-1},\dots,e^{i\alpha_{N-2}}\sin(\theta)/\sqrt{N-1}\}.
\end{eqnarray}
Since $U(\theta)$ is unitary, the inner product between these two rows must be $0$, i.e.,
\begin{equation}\label{VS}
2k+e^{i\alpha_1}+\dots+e^{i\alpha_{N-2}}=0,
\end{equation}
where
\begin{equation}\label{B5}
k=\cot(\theta)\sqrt{N-1}.
\end{equation}
It is known that a sum of $N'$ terms of the form $e^{i\alpha_1}+\dots+e^{i\alpha_{N'}}$, composed by $m$-th roots of unity with $m=p_1^{a_1}\dots p_r^{a_r}$, can vanish only if there exists numbers $x_j\in\mathbb{N}_0$ such that $N'=x_1p_1+...+x_rp_r$, where $p_1,\dots p_r$ are prime numbers and $a_1,\dots,a_r\in\mathbb{N}$ \cite{LL00}. Since $2k=1+1+\cdots+1$, equation (\ref{VS}) can be understood as a vanishing sum of $N'=2k+N-2$ unimodular numbers. Therefore, a necessary condition for the existence of a Gram matrix composed by $m$-th roots of unity with $m=p_1^{a_1}\dots p_r^{a_r}$ is the existence of  numbers $x_j\in\mathbb{N}_0$ such that $2k+N-2=x_1p_1+\dots+x_rp_r$.
\end{proof}

\end{document}